\documentclass[11pt,tightenlines]{revtex4}
\usepackage{amsfonts}
\usepackage{amsmath}
\usepackage{txfonts}
\usepackage{graphicx}
\usepackage{amssymb}
\usepackage{color}
\usepackage{wrapfig,lipsum,booktabs}
\usepackage{floatrow}
\usepackage{epstopdf}
\usepackage{hyperref}
\usepackage[normalem]{ulem}

\newcommand{\beginsupplement}{
        \setcounter{table}{0}
        \renewcommand{\thetable}{S\arabic{table}}
        \setcounter{figure}{0}
        \renewcommand{\thefigure}{S\arabic{figure}}
     }
     
\begin{document}
\title{Topological mapping of space in bat hippocampus}
\author{Kentaro Hoffman$^{1}$, Andrey Babichev$^{1,2}$ and Yuri Dabaghian$^{1,2}$}
\affiliation{$^1$Jan and Dan Duncan Neurological Research Institute, Baylor College of Medicine,
Houston, TX 77030, \\
$^2$Department of Computational and Applied Mathematics, 
Rice University, Houston, TX 77005}

\date{\today}
\begin{abstract}
Mammalian hippocampus plays a key role in spatial learning and memory, but the exact nature of the 
hippocampal representation of space is still being explored. Recently, there has been a fair amount of 
success in modeling hippocampal spatial maps in rats, assuming a topological perspective on spatial 
information processing. In this paper, we use the topological model to study $3D$ learning in bats, which 
produces several insights into neurophysiological mechanisms of the hippocampal spatial mapping. 
First, we demonstrate functional importance of the cell assemblies for producing accurate maps of 
the $3D$ environments. Second, the model suggests that the readout neurons in these cell assemblies 
should function as integrators of synaptic inputs, rather than detectors of place cells' coactivity and 
allows estimating the integration time window. Lastly, the model suggests that, in contrast with relatively 
slow moving rats, suppressing $\theta$-precession in bats improves the place cells capacity to encode 
spatial maps, which is consistent with the experimental observations. 
\end{abstract}
\maketitle

\newpage

\section{Introduction}
\label{section:intro}
The principal neurons in mammals' hippocampus---the place cells---fire in discrete locations within the 
environment---their respective place fields \cite{OKeefe1,Best}. The spatial layout of the place fields---the place 
field map---is commonly viewed as a representation of the animal's cognitive map of space, although the exact 
link between them remains unclear \cite{OKeefe2}. Experiments in ''morphing'' $2D$ environments demonstrate 
that the place field maps recorded in rats are ``flexible'': as the environment is deformed, the place fields change 
their shapes, sizes and locations, while preserving their relative positions \cite{Gothard,Leutgeb,Wills,Touretzky}, 
which suggests that the sequential order of place cells' (co)activity induced by the animal's moves through the place 
fields remains invariant within a certain range of geometric transformations. Moreover, the temporal patterns of place 
cell coactivity is preserved \cite{Diba, eLife}, which implies that the place cells' spiking encodes a coarse framework 
of qualitative spatiotemporal relationships, i.e., that the hippocampal map is topological in nature \cite{eLife,Alvernhe,Poucet,Wu}.

Recently, there appeared a few topological models of the hippocampal map \cite{Chen1, Chen2, Curto}. In particular, 
the approach proposed in \cite{Dabaghian,Arai} allows integrating the local spatial information provided by the individual 
place cells into a large-scale topological representation of the environment. The idea of such integration is based on the 
\v{C}ech's theorem, according to which the pattern of overlaps between regular spatial domains---the regions---$U_1$, 
$U_2$, … $U_N$, covering a space $X$, encodes the topological structure of $X$ \cite{Hatcher}. Specifically, the covering 
regions are used to construct the nerve of the cover, $\mathcal{N}$---a simplicial complex whose 0D vertices correspond 
to the covering regions $U_i$, the $1D$ links---to pairwise overlaps $U_i \cap U_j$, the $2D$ facets---to triple overlaps 
$U_i \cap U_i \cap U_i$ and so forth. The \v{C}ech's theorem ascertains that if all of the overlaps between the $U_i$s are 
contractible in $X$, then $\mathcal{N}$ is topologically equivalent to $X$. An  implication is that, if the place fields cover the 
environment sufficiently densely, then their overlaps should encode its topology. Moreover, since these overlaps are 
represented by the place cells' coactivities, a similar construction can be carried out in temporal domain \cite{Curto,Dabaghian,Arai}:
if the animal enters a location in which several place fields overlap, the corresponding place cells produce (with a certain probability) 
temporally overlapping spike trains, which can be received and processed by the downstream brain areas (Suppl.  Fig. 1). 
In other words, place cells' spiking encodes a ``temporal'' analogue of $\mathcal{N}$---a coactivity complex, $\mathcal{T}$, 
the vertices of which correspond to active place cells $c_i$, $1D$ links---to pairs of coactive cells $[c_i, c_j]$, $2D$ facets---to 
coactive triples $[c_i, c_j, c_k]$, etc. 

\begin{figure} 
\includegraphics[scale=0.84]{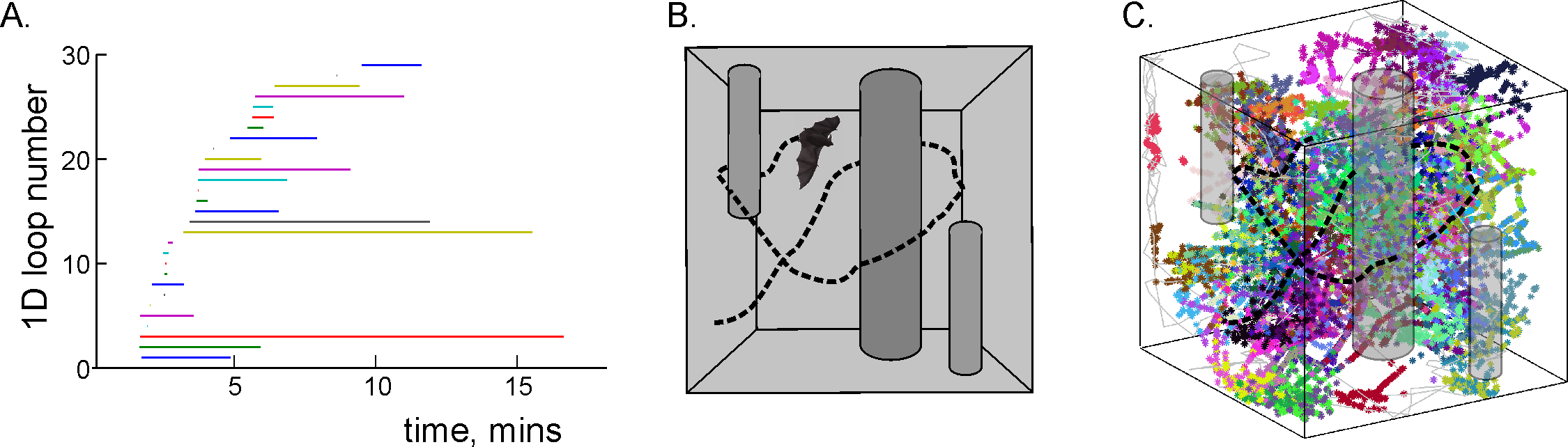}
\caption{\label{PFs} \textbf{Topological map of the bat's environment}. (\textbf{A}) Timelines of $1D$ loops (colored 
horizontal lines) encoded in a coactivity complex $\mathcal{T}$. For as long as a given loop $1D$ persists, it 
indicates a noncontractible hole in $\mathcal{T}$. In this case, the $1D$ topological loops begin to appear (and 
hence the holes start to form) in about 2 minutes, and disappear in about 17 minutes, when all the holes in 
$\mathcal{T}$ close up. (\textbf{B}) A view into a simulated $3D$ environment ($290 \times 280 \times 270$ cm, sizes 
taken from \cite{Ulanovsky, Yartsev}) that contains one vertical column and two protrusions---a stalactite hanging 
over 50 cm from the ceiling, and a 50 cm tall stalagmite. A portion of the simulated trajectory is shown by dashed 
line. (\textbf{C}) Simulated spikes produced by a virtual bat form $3D$ spatial clusters---the place fields. Spikes 
produced by different cells are marked by different colors.} 
\end{figure}

By construction, the coactivity complex incorporates, at any given moment of time $t$, the entire pool of coactivities 
produced by the place cell ensemble. Hence, it provides a framework for representing spatial information encoded 
by the place cells. For example, a sequence of the place cell combinations ignited along a particular path $\gamma$ 
corresponds to a sequence of ``coactivity simplexes''---a simplicial path $\Gamma$ that represents $\gamma$ in 
$\mathcal{T}$. It was shown in \cite{Dabaghian,Arai} that if the coactivity complex is sufficiently large (i.e., includes 
a sufficient number of the coactivity events) and if the parameters of the place cell activity fall into the biological range, 
then $\mathcal{T}$ correctly captures the topology of the physical environment, $\mathcal{E}$. For example, a 
non-contractible simplicial path corresponds to a class of the physical paths that enclose unreachable or yet unexplored 
parts of the environment. Similarly, two topologically equivalent simplicial paths $\Gamma_1 \sim \Gamma_2$ in $\mathcal{T}$ 
represent physical paths $\gamma_1$ and $\gamma_2$ that can be deformed into one another. However, such information 
is not produced immediately: as the animal begins to navigate a new environment, the coactivity complex $\mathcal{T}$ 
is small and provides an incomplete and fragmented representation of space; moreover, an undeveloped complex 
$\mathcal{T}$ typically consists of several disjoint components, each one of which may contain gaps and holes that do 
not correspond to physically inaccessible spatial domains. As the animal continues to navigate, more combinations of 
coactive place cells are detected, and the ``spurious'' gaps disappear, yielding a coactivity complex that faithfully represents 
the topological structure of the environment. Thus, the progress of spatial learning can be quantified in terms of the 
evolving coactivity complex's structure, e.g., a list of the holes that it contains at a given moment of time $t$, or a list of 
surfaces---loops---that encapsulate these holes \cite{Dabaghian, Hatcher}. Both the holes and the loops are counted up 
to topological equivalence. For example, if two such loops can be deformed into one another, then they are counted as different 
shapes of the same topological object---a topological loop. The number of inequivalent topological loops of a dimensionality 
$n$ is known as the $n$th Betti number, $b_n$, and the list of all Betti numbers, $(b_0, b_1, b_2, ...)$, provides a convenient 
``barcode'' of the space's topological structure \cite{Ghrist}. Methods of the Persistent Homology theory allow identifying the 
topological loops in the hippocampal representation of the environment and comparing the resulting Betti numbers 
$b_n(\mathcal{T})$ with the Betti numbers of the underlying environment, $b_n(\mathcal{E})$. In addition, it is possible to 
compute the minimal time, $T_{\min}$, after which the low-dimensional topological structure of $\mathcal{T}$ matches 
the topology of the environment, which serves as a theoretical estimate of the time required to learn a given space 
(Figure~\ref{PFs}C). 

Below we apply this model to investigate spatial learning in bats that learn $3D$ representations of their environments 
\cite{Ulanovsky, Yartsev}, which produces a number of neurophysiological insights. 
Specifically, we simulated place cell spiking activity in a bat navigating a small cave with one vertical column and two 
vertical protrusions, representing a stalactite and a stalagmite (Figure~\ref{PFs}A). The topological barcode of this 
environment is the same as in the $2D$ open field studied in \cite{Dabaghian,Arai}: $b_0 = 1$ (the cave is connected), 
$b_1 = 1$ (one $1D$ topological loop represents paths encircling the column, but not the contractible stalactite and stalagmite) 
and $b_{n > 1} = 0$ (no loops in higher dimensions). However, the increased complexity of the space mapping task 
elucidates several neurophysiological properties of the hippocampal network that were not explicitly addressed in the 
previous models.

\section{Results}
\label{section:results}

\textbf{Simplicial coactivity complexes}. Our first observation was that, in contrast with the $2D$ environment in which the 
correct topological signature emerged in a matter of minutes, the simulated place cell coactivity in $3D$ failed to represent 
the cave's topology. In particular, the first Betti number of the coactivity complex often vanished, $b_1(\mathcal{T}) = 0$, 
i.e., the place cells did not capture the most salient feature of the navigated space---the central column, although the bat's 
trajectory encircled it multiple times.

We reasoned that the discrepancy between $b_1(\mathcal{T})$ and $b_1(\mathcal{E})$ was due to the relatively high 
speed of the bat's movements, which caused a mismatch between the temporal pattern of place cell coactivities and the 
spatial pattern of the place field overlaps. Indeed, if the bat is flying at the speed $v$, then place cells cofire within a 
coactivity time window $w$, even if their place fields are up to $d \approx vw$ apart. If detected by the downstream neurons, 
these coactivities may lead to an inaccurate representation of space by linking the place cell representations of the physically 
separated spatial domains. In our simulations, the speed $v$ of the bat reached at times 2 m/sec, while the place cell inputs 
were integrated over the coactivity window $w = 0.25$ sec \cite{Arai, Mizuseki, Huhn}. As a result, place cell coactivities 
could falsely encode overlaps between place fields that are physically up to 50 cm apart. In particular, place fields across the 
column may appear ``connected,'' in which case the column will fail to produce a hole in $\mathcal{T}$.

\textbf{Cell assembly constraints}. The result outlined above suggests that the pool of place cell coactivities must be additionally 
constrained to prevent the appearances of ``faulty'' connections, which, in fact, appeals to well-known neurophysiological 
phenomenon observed in the hippocampal network. Electrophysiological studies suggest that certain select groups of place 
cells form functionally interconnected assemblies, which drive their respective ``reader-classifier'' or ``readout'' neurons in 
the downstream networks \cite{Harris, Buzsaki1}. Spiking of the readout neurons ``actualizes'' the information provided by 
the place cells: if the coactivity of a place cell assembly does not elicit a response of a readout neuron, the corresponding 
connectivity information does not contribute to the hippocampal map \cite{Buzsaki1}. Thus, a particular selection of the 
admissible place cell combinations is determined not only by the place coactivities but also by the architecture of the 
hippocampal cell assembly network, which constrains the pool of place cell coactivities. 

It is widely believed that the readout neurons function as ``all or none'' coactivity detectors, i.e., that they respond to nearly 
simultaneous activity of the presynaptic cells \cite{Buzsaki1}. Curiously, the topological approach based on the \v{C}ech's theorem 
corroborates with this point of view: the fact that the nerve complex $\mathcal{N}$ is derived from the spatial overlaps 
between the regions suggests that a readout neuron should identify these overlaps by detecting nearly simultaneous place cell 
activity at the times when the animal visits them. 

\begin{figure} 
\includegraphics[scale=0.84]{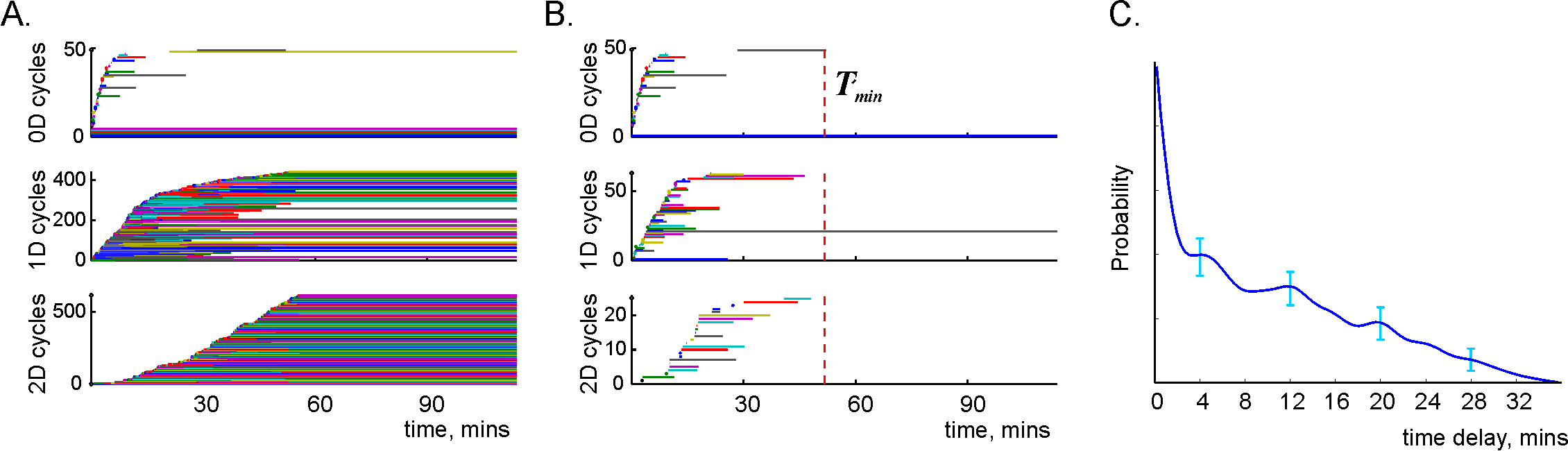}
\caption{\label{ThOn} \textbf{Learning a topological map by detecting place cell coactivity}. (\textbf{A}) A simplicial coactivity 
complex, built by detecting high order coactivity events, produces large numbers of persisting topological loops in low 
dimensions. In particular, there are five persistent 0D topological loops (top panel), which implies that hippocampal map 
of the environment is fragmented in pieces. In addition, there are many persistent loops in $1D$ (middle panel) which 
implies that the coactivity complex encodes many noncontractible paths, whereas there is only one class of noncontractible 
loops in the physical environment. (\textbf{B}) The number of topological loops in the clique coactivity complex is much smaller. 
In fact, after the $T_{\min}$ = 28 minutes (red vertical dashed line), most topological loops disappear, leaving only one 
topological loop in 0D (which correctly represents the caves connectivity) and one loop in $1D$, which represents physical 
paths circling around the central column. The loops in higher dimensions contract (only the $2D$ loops are shown, bottom 
panel), and thus the correct topological barcode ($b_0 = 1$, $b_1 = 1$, $b_{n > 1} = 0$) emerges. (\textbf{C}) A particular 
connection may be identified in two ways: either as a clique (completing at $t_c$, the moment when all the pairs have been 
accumulated), or simultaneously, as a simplex all cells of which are observed at the moment $t_s$. The figure shows the 
probability distribution of the differences between the first appearance times, $\Delta =  t_c - t_s$. Since all $\Delta$s 
are positive, cliques always appear sooner than simplexes.} 
\end{figure} 

These considerations suggest a simple phenomenological solution to the constraint selection task: to exclude the ``spurious'' 
connections, we built the coactivity complex $\mathcal{T}$, using only the coactivities produced by place cells with the overlapping 
place fields. However, our subsequent simulations revealed that the cell assembly constraint is too restrictive, because the resulting 
coactivity complex broke into pieces (on average, $b_0 = 2.3$) and produced a large number of topological loops in $1D$ and 
$2D$ (on average, $b_1 = 36$ and $b_2 = 410$, Figure~\ref{ThOn}A).

From a biological perspective, this implies that the hippocampal map of the environment remained fragmented and riddled with 
holes, most of which do not correspond to the actual topological obstacles encountered by the bat. In other words, the simulated 
cell assembly network, wired to detect place cells coactivities, fails to learn the correct path connectivity of space, which suggests 
that the system may employ an alternative mechanism of reading out the coactivity information. 

\textbf{Clique coactivity complexes}. From a mathematical perspective, the higher order overlaps between regions can be not only 
empirically detected, but also derived from the lower order overlaps. According to Helly's theorem \cite{Avis}, a collection of 
$N > D +1$ convex regions in $D$-dimensional Euclidean space will necessarily have a nonempty common intersection if every 
$D +1$ of them intersect. For example, a collection of $N \geq 4$ planar regions has a common intersection if any three of them 
overlap and a collection of $N \geq 5$ regions have a common intersection in $3D$ if any four of them overlap (Suppl.  Fig. 2). 
From the perspective of the \v{C}ech's theorem, this suggests that high-dimensional simplexes in $\mathcal{N}$ can be deduced 
from their low-dimensional simplexes, which opens new possibilities for constructing coactivity complexes.

Numerical simulations demonstrate that, in fact, the approach of Helly's theorem can be extended beyond its strict mathematical 
validity. For example, high order overlaps between place fields in $2D$ environments can be reliably identified by detecting 
graph-theoretic cliques of pairwise overlaps between them (Suppl.  Fig. 2A,B). 

Moreover, this information is captured by the place cell coactivity: in the case of the triple place field overlaps, the three pair 
coactivities, $[c_i,c_j]$, $[c_j, c_k]$ and $[c_i, c_k]$, mark an existing triple overlap of the corresponding place fields in 
over 90$\%$ of cases, and for the higher order overlaps this percent is even higher. This implies that most cliques in the 
coactivity graph $G$, defined by the connectivity matrix 
\begin{equation}
C_{ij} = \begin{cases} 1 &\mbox{if cells } c_i,c_j  \mbox{are coactive and their respective place fields overlap} \\ 
0 & \mbox{otherwise}. 
\end{cases}
\label{C}
\end{equation}
correspond to the simplexes of the restricted coactivity complex $\mathcal{T}$, i.e., that the topological structure of the 
``clique complex,'' $\mathcal{T}_{cq}$, can approximate the topological structure of $\mathcal{T}$. 

In \cite{Babichev} we demonstrated that in $2D$, the clique coactivity complexes do not only capture the topology of the 
environment, but often perform better than simplicial coactivity complexes. Similar effects are also observed in the $3D$ case.
In our simulations, the number of topological loops in $\mathcal{T}_{cq}$ was much smaller than the number of loops in 
$\mathcal{T}$. Moreover, $\mathcal{T}_{cq}$ produced the correct topological signature on average in about 
$T_{\min}$ = 28 minutes---a biologically plausible period of time (Figure~\ref{ThOn}B).

The success of the clique coactivity complex has a simple intuitive explanation. First, the pairwise coactivities of the place 
cells are produced when the animal enters the domains where at least two place fields overlap. Since these domains are
bigger than the domains of the higher order overlaps (Suppl. Fig. 2C), the pairwise coactivities are produced and detected 
more reliably than the high-order coactivities. Second, the process of detecting the pairwise coactivities is spread over time. 
For example, in order to identify a third order coactivity clique, $[c_i, c_j, c_k]$, one can first detect the coactive pair 
$[c_i, c_j]$, then the pair $[c_j, c_k]$, and then $[c_i, c_k]$, whereas in order to encode a coactivity simplex 
$[c_i, c_j, c_k]$, all three cells must become active within the same coactivity window $w$. As shown on Figure~\ref{ThOn}C, 
the higher order combinations take longer to appear than the matching collection of pairs---in fact, most coactivity cliques 
that can be ``assembled'' over an extended observation process, are never observed as simultaneous coactivity events 
(Suppl. Fig. 3). This implies that the clique coactivity complex is typically bigger and forms faster, and hence the transient 
topological loops in $\mathcal{T}_{cq}$ contract sooner. 

\begin{figure} 
\includegraphics[scale=0.84]{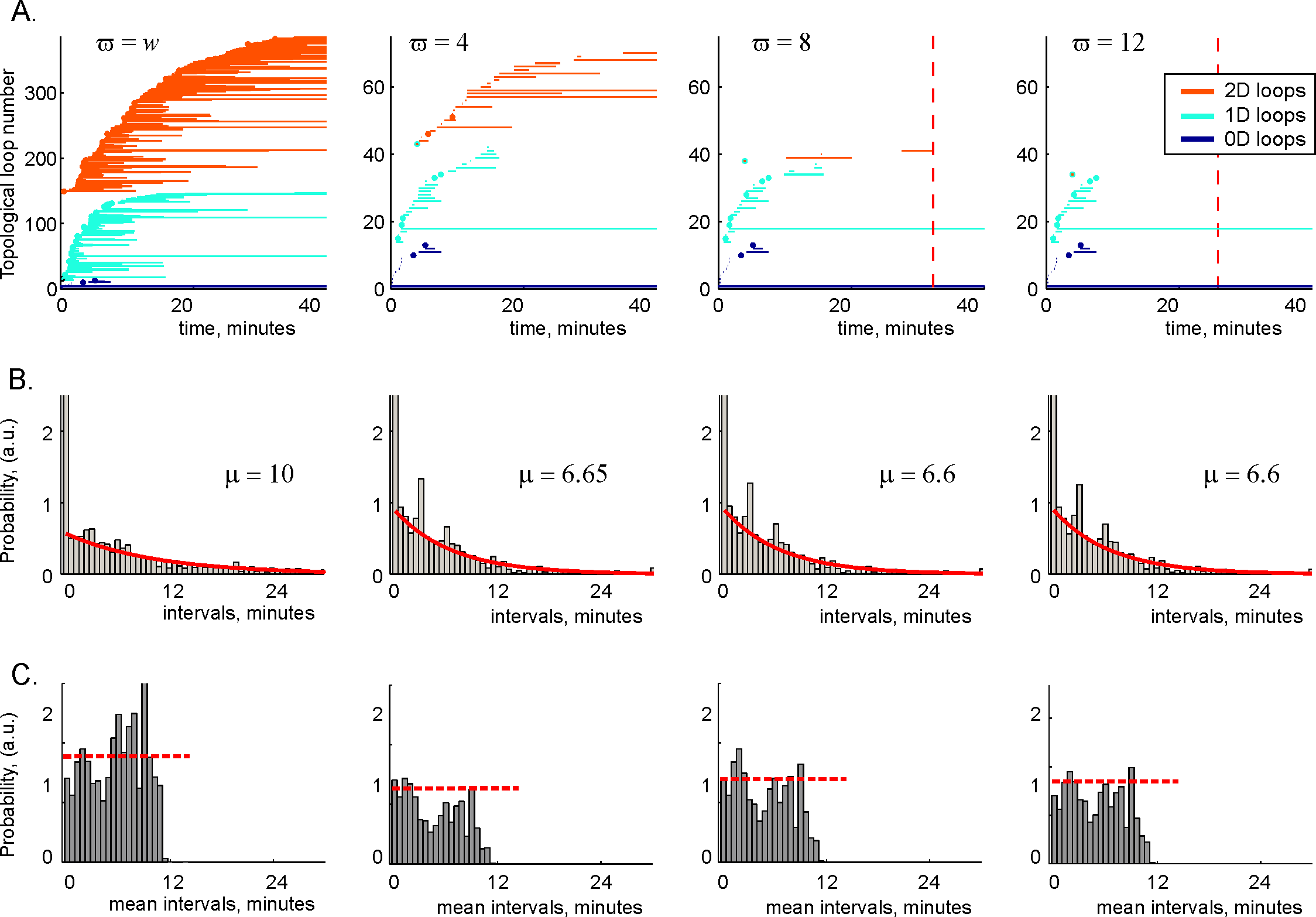}
\caption{\label{Wind} \textbf{Timelines of the topological loops in the restricted clique simplicial complexes, obtained for different 
dendritic integration times.} (\textbf{A}) For $\varpi = w$ (in which case the clique complex reduces to the simplicial coactivity 
complex) spatial learning fails due to numerous “spurious” topological loops that persist indefinitely in 0D, $1D$ and in $2D$. 
However, as the dendritic integration time increases to $\varpi$ = 4 min, the number of topological loops drops, and only 
occasional spurious loops remain for $\varpi$ = 8 minutes. At $\varpi$ = 12 minutes the last $2D$ loops disappear, and 
only one persistent loop remains in $1D$ and 0D after 28 mins, indicating emergence of the correct topological signature. 
(\textbf{B}) The distribution of the time intervals between the pairs impinging on the readout neurons (see formula (2)). 
(\textbf{C}) Distribution of the same intervals, averaged for each clique.} 
\end{figure} 

\textbf{Integration times}. From a physiological perspective, the qualitative difference between the results produced by the 
``simplicial'' and the ``clique'' approaches to counting the place cell coactivities suggests that the readout networks should 
build their spiking responses not by detecting rare high-order events, but by integrating low order coactivity inputs. The 
physiological mechanism for such integration may be based on complex subthreshold summation of the action potentials 
impinging on the dendritic tree of the readout neuron. Once a sufficient number of low order coactivity inputs has been 
received, the readout neuron may produce an action potential, thus actualizing the information about the $n$th order 
connection between the regions encoded by the place cells. 

Although our modeling approach does not directly address spike integration mechanisms, it allows optimizing parameters 
in a particular readout algorithm, based on the frequency of the place cell (co)activations produced by the animal's movements
through the environment. For example, the model predicts that the coactivity window used by the coincidence detector 
readout neurons in $2D$ environments should be about $w \sim 200$ msec wide (smaller $w$s lead to a rapid increase of the 
learning times and larger $w$s lead to instability of learning \cite{Arai}), which falls into the physiological range of values 
\cite{Mizuseki, Huhn}. 

What would be then the model's estimate of the time required by the ``integrator'' neurons to accumulate the place cell 
coactivities---the clique integration time, $\varpi$---to produce a complex that reliably represents space? On the one hand, 
longer clique integration windows allow collecting more coactivities for assembling the cliques, which improves the structure 
of the coactivity complex (compare Figure~\ref{ThOn}B and Figure~\ref{ThOn}A). 
On the other hand, the larger the $\varpi$ is, the longer the intervals between the consecutive coactivity inputs $\tau$ can be, 
so the information about the presynaptic inputs has to be retained for longer. In contrast, the smaller is $\varpi$, the tighter the 
coactivities are ``packed'' in the integration window. In cases when $\varpi \approx w$, the integrator neuron turns into a 
coincidence detector, and the clique coactivity complex reduces to the simplicial coactivity complex (i.e., when $\varpi \approx w$, 
the results shown on Figure~\ref{ThOn}B reduce to Figure~\ref{ThOn}A). By testing several values of $\varpi$, we found that 
the place cell map converges if $\varpi \gtrsim 8$ minutes (Figure~\ref{Wind}A), making this value the model's estimate of 
the clique integration window. Physiologically, this value corresponds naturally to the timescale of working memory (the memory 
functions responsible for the transient holding, processing and retaining the partial results of learning and memory updating 
\cite{Goldman}).

We evaluated the combined statistics of the time intervals, $\tau$, between the pairs of action potentials received by the 
readout neurons within an integration window, across the cell assembly network. As shown on (Figure~\ref{Wind}B), the 
distribution of these intervals consists of two parts: a sharp delta-like peak, $\Delta(\tau) $, concentrated at small values 
of $\tau$, and an exponential tail,
\begin{equation}
P(\tau) = C_1\Delta(\tau) +C_2e^{-\mu\tau}.
\label{tau}
\end{equation}
The exponential rate $\mu$ is higher for short $\varpi$s and decreases towards a stable value of about $\mu \approx 6.6$ 
minutes as $\varpi$ grows. The fact that building a typical clique requires detecting several coactivities (e.g, a fourth order 
clique contains six parwise connections) while mean interval $\mu$ between coactivities is comparable to the size of the 
integration window $\varpi$, suggests that most coactive inputs are detected ``on the spot,'' within a small number of 
consecutive coactivity windows, at the time when the bat crosses a domain where several place fields overlap. 
This contribution is described by the first part of the distribution, $\Delta(\tau)$, whereas the second part---the exponential 
tail---represents the connections accumulated over time. 

To better understand the statistics of the inputs provided by the cells within the individual cell assemblies we computed, for 
each $\varpi$, the mean inter-activity interval, $\bar \tau_{\sigma}$, for each clique $\sigma$ and studied the statistics of 
these values. The results shown on (Figure~\ref{Wind}C) suggest that the mean inter-activity intervals are distributed much 
more uniformly, i.e., the population of cell assemblies in which the coactivity integration occurs rapidly is about the same as 
the population of assemblies that take intermediate or longer times to integrate inputs.

\begin{figure} 
\includegraphics[scale=0.84]{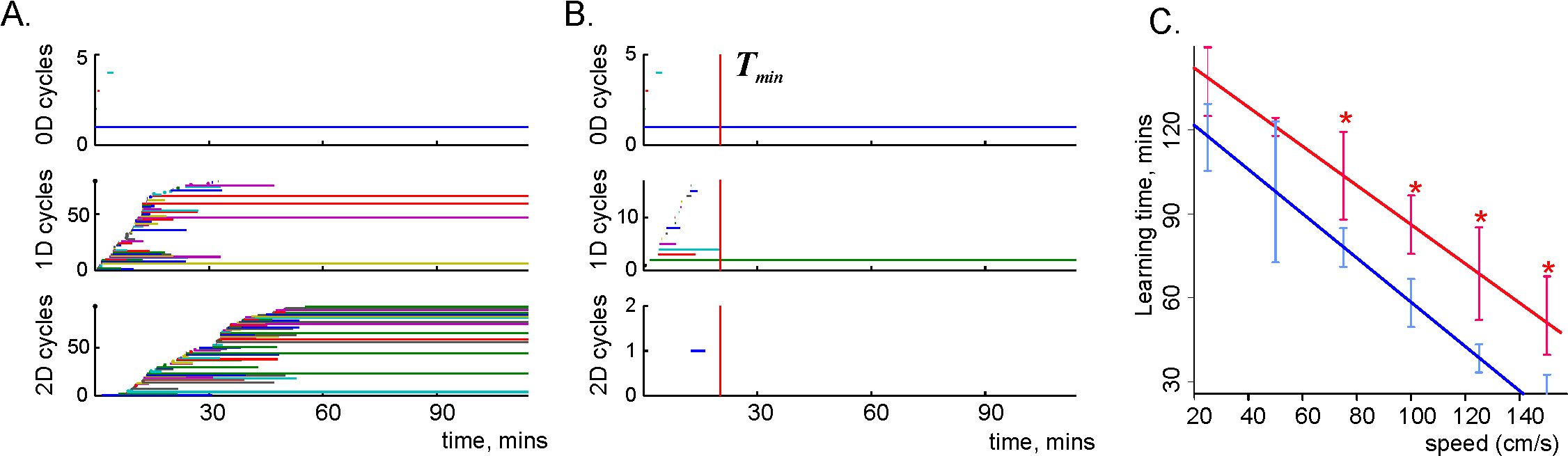}
\caption{\label{ThOff} \textbf{Suppression of the $\theta$-phase precession improves learning.} (\textbf{A}) If the 
$\theta$-precession in the place cells producing the simplicial coactivity complex is supressed, then the number of the 
topological loops in it drops significanlty in all dimensions (compare to Figure~\ref{ThOn}A,B in which $\theta$-precession 
is present). (\textbf{B}) Similar effect is observed in the clique coactivity complex, if the place cells do not $\theta$-precess. 
Moreover the learning time $T_{\min}$ = 18 minutes is smaller than in the $\theta$-precessing place cell ensembles. 
(\textbf{C}) The average time it took the coactivity complex, $\mathcal{T}$ to converge to a stable signature, as a function 
of the bat's mean speed. The times computed for the $\theta$-driven cell assemblies are shown in red and those without 
$\theta$-precession are colored blue, statistically significant differences are marked by the asterisks over the error bars.} 
\end{figure} 

\textbf{Theta precession} is a coupling between the timing of neuronal spikes and the phase of the theta ($\theta$, 4-12 Hz) 
component of the extracellular field potential, observed in the rat's hippocampus. As the rat proceeds through a place field, 
the corresponding place cell spikes at progressively earlier phases of $\theta$-oscillations \cite{Skaggs, Buzsaki2}. 
This phenomenon produces a strong effect on rats' spatial learning \cite{Buzsaki3}: suppressing $\theta$-precession with 
cannabinoids correlates with reduced ability to solve spatial tasks and poorer spatial memory \cite{Robbe1, Robbe2}. 
In contrast, $\theta$-precession in bats is not clearly manifested: according to \cite{Heys}, less than 4$\%$ of the bat's place 
cells exhibit significant $\theta$-modulated firing.

From a biological perspective, the functional importance of $\theta$-precession for rats navigating $2D$ environments and 
lack of thereof in bats navigating $3D$ spaces might suggest that in the latter case, $\theta$-precession may not produce a 
similar positive effect on spatial learning. To test this hypothesis, we numerically suppressed the $\theta$-precession in the 
simulated place cells, which lead to a noticeable decrease of the number of spurious topological loops both in the simplicial 
and in the clique coactivity complexes (Figure~\ref{ThOff}). Second, the learning time produced by the clique complex, 
$T_{\min}$ = 18 minutes, became shorter than in the $\theta$-driven coactivity complex.

To explain these results, we reasoned as follows. One can view the effect of $\theta$-precession from two perspectives: on 
the one hand, it synchronizes place cells' spiking and hence increases their coactivity rate \cite{Skaggs, Buzsaki2}, which 
speeds up map learning. On the other hand, the fact that the place cells can spike only at specific phases of $\theta$ can be 
viewed as a constraint that reduces the probability of cells' spiking at every given moment. In \cite{Arai} we demonstrated 
that in relatively slow moving rats, the probability of producing spikes during a typical passage through the place field is 
sufficiently high despite the $\theta$-constraint, so that the main effect of $\theta$-oscillations is spike-synchronization. In 
contrast, the results of Figure~\ref{ThOff} suggest that in bats, the place cells have time to produce only a few spikes during 
a high speed flight through the $3D$ place fields, so that the $\theta$-precession only further reduces the cells' chances of 
being (co)active. 

We tested these results by simulating bat's movements at different mean speeds: (25, 50, 75, 100, 125, and 150 cm/sec). 
For each case, 10 simulations were conducted using $\theta$-precessing place cells and 10 simulations with the $\theta$-
precession turned off. The results on the Figure~\ref{ThOff}C show that at no point $\theta$-on ensembles learned faster 
than the $\theta$-off ensembles: average time until a stable signature was reached in the $\theta$-off case was about 
30$\%$ less than the time required by the $\theta$-modulated place cell ensembles.

\section{Discussion}
\label{section:discussion}

The topological model of the place cell map of space proposed in \cite{Dabaghian,Arai} provides a framework for bringing 
together different scales of spatial learning: the macroscale, i.e., the topological and geometrical parameters of the encoded 
loops, paths, holes, and the microscale, i.e., the parameters of the hippocampal network's neurophysiology. Applying this 
model to describe a bat navigating a $3D$ space reveals several interrelationships between these scales that are implicit 
in the $2D$ case, but which have a number of important physiological implications.

First, the failure of the na\"{i}ve counting of the place cell cofirings indicates a functional necessity of ``thinning out'' the pool of 
coactivities using a cell assembly network. Qualitatively, this result is based on simple observation: if the readout neurons 
respond unrestrictedly to the place cell coactivities then the animal's rapid moves across the environment will necessarily 
encode false connections between remote spatial regions, which will result in an incorrect map of space. Second, a striking 
difference between the simplicial and the clique complexes' ability to capture the topology of the $3D$ environment suggests 
that the information about the high order spatial relationships should be deduced from low order coactivity events, rather 
than instantaneously detected. The qualitative reasons for this effect are also transparent: the high order coactivity events, 
represented by the simplexes, are rarer and harder to detect than the matching collections of pairwise coactivities, represented 
by the cliques. Third, the amount of time over which readout neurons integrate inputs should be longer than observed in previous 
computational or in vitro studies, in which neurons were studied individually and independently from the task solved by the net 
neuronal ensemble \cite{London, Brody, Magee, Spruston, Rall, Jarsky}. In contrast, our approach provides a basic contextual 
description of neuronal activity: the frequency with which the place cells' action potentials impinge on the readout neuron 
depends on the frequency of the animal's visits to specific spatial locations. This frequency would not change significantly if the 
number of simulated place cells would increase, which suggests that our estimates of the ``clique integration windows'' $\varpi$ 
may give a correct qualitative estimate of the physiological dendritic integration times required by the downstream networks. 

Lastly, the model provides a functional insight into why the $\theta$-precession is physiologically suppressed in bats' hippocampi. 
Thus, our model provides an example of a ``top-down'' approach, in which neurophysiological properties of the network are 
deduced from the task solved by the network.

\section{Methods}
\label{section:methods}
\textbf{The navigational parameters}: the mean speed of the bat ($v_{\rm{mean}} = 66$ cm/s, $v_{\max} = 150$ cm/s), 
and the dimensions of the environment ($290 \times 280 \times 270$ cm) were taken from \cite{Yartsev}. 
For increased reliability of the results, the duration of the navigation session (120 min) was longer than reported in 
\cite{Yartsev}.

\textbf{Place fields}. The centers of the place fields, $r_c = (x_c, y_c, z_c)$, were uniformly distributed over the 
environment, to simulate non-preferential representation of locations. Given the typical sizes of the place fields 
($L_c = 95$ cm), the size of the simulated place cell ensemble was chosen to create enough $2D$ and $3D$ simplexes 
in the coactivity complex, sufficient to build a $3D$ map. Computationally, we could afford about seven place fields per 
dimension, i.e., $N_c = 343$ place cells total, which corresponds to about $N_c = 50$ cells in a the planar $1 \times 1$ m 
environment studied in \cite{Dabaghian,Arai}. The Poisson spiking rate of a place cell $c$ at a point $r(t) = (x(t), y(t), z(t))$ is 
\begin{equation}
\lambda_c(r)=f_c e^{-\frac{(r-r_c)^2}{2s_c^2}}.
\label{lambda}
\end{equation}
The peak firing rates $f_c$ of different cells were log-normally distributed around the typical  experimental value, 
$f = 8$ Hz and $s_c = L_c /3$ \cite{Yartsev}. 

\textbf{$\theta$-phase precession}. As the animal enters the field of a cell $c$, and moves over a distance $l$ towards the center, 
the preferred spiking phase is $\varphi_{\theta,c} \approx 2\pi (1-l/L_c)$ \cite{Buzsaki2,Huxter}. To simulate the coupling 
between the firing rate and the $\theta$-phase, we modulated the original Gaussian firing rate by a $\theta$-factor 
$\Lambda_{\theta,c}(\varphi)$, 
\begin{equation}
\Lambda_{\theta,c}=e^{-\frac{(\varphi-\varphi_{\theta,c})^2}{2\varepsilon_c^2}}.
\label{lambdaTheta}
\end{equation}
The width, $\varepsilon$, of the Gaussian was defined in \cite{Arai}, as the ratio of the mean distance that the animal travels 
during one $\theta$-cycle to the size of the place field, $\varepsilon  = 2\pi v /L\omega_{\theta}$, where $v$ is the rat's 
speed and $\omega_{\theta}/2\pi$ is the frequency of the $\theta$-signal.

\textbf{Cell assembly constraint}. The functional connectivity between place cells in the hippocampal network is described 
by a graph $G$, the vertices of which correspond to the active place cells and links to coactive pairs, restricted by the 
constraints (see formula (1)). The connectivity matrix $A_{ij}$ of $G$ is defined as follows. First, we define the relational 
matrix $C_{ij}$: $C_{ij} = 1$ if the cells $c_i$ and $c_j$ exhibit coactivity during the navigation period and $C_{ij} = 0$ 
otherwise. The place field map's connectivity matrix, $P_{ij}$, is based on the place fields' spatial overlap: $P_{ij} = 1$ 
if the distance between the place field centers is smaller than the sum of their half-sizes, $d(c_i,c_j) \leq (L_i + L_j)/2$, 
and $P_{ij} = 0$ otherwise. In principle, the matrix $P_{ij}$ can be deduced directly from the temporal pattern of place 
cell coactivity \cite{Babichev}; however we used the place field information to simplify our analyses. To constrain the pool 
of coactivities by the place field map structure, we used the Hadamard product, $A_{ij} = C_{ij} P_{ij}$. The simplexes 
of the clique coactivity complex, $\mathcal{T}_{cq}$, correspond to ``cliques'' of the relational graph $G$.

\textbf{Coactivity}. A group of cells $c_1$, $c_2$, ..., $c_n$, counts as coactive, if each one of them fires at least two spikes 
within the coactivity window $w \approx 250$ msec, i.e., in less than two $\theta$-periods. In \cite{Arai} we demonstrated 
this value of $w$ is optimal both in presence and in absence of the $\theta$-precession.

\section{Acknowledgments}
\label{section:acknow}

The work was supported in part by Houston Bioinformatics Endowment Fund, the W. M. 
Keck Foundation grant for pioneering research and by the NSF 1422438 grant.

\newpage
\beginsupplement

\section{Supplementary Figures}
\label{section:SupplFigs}

\begin{figure}[ht] 
\includegraphics[scale=0.84]{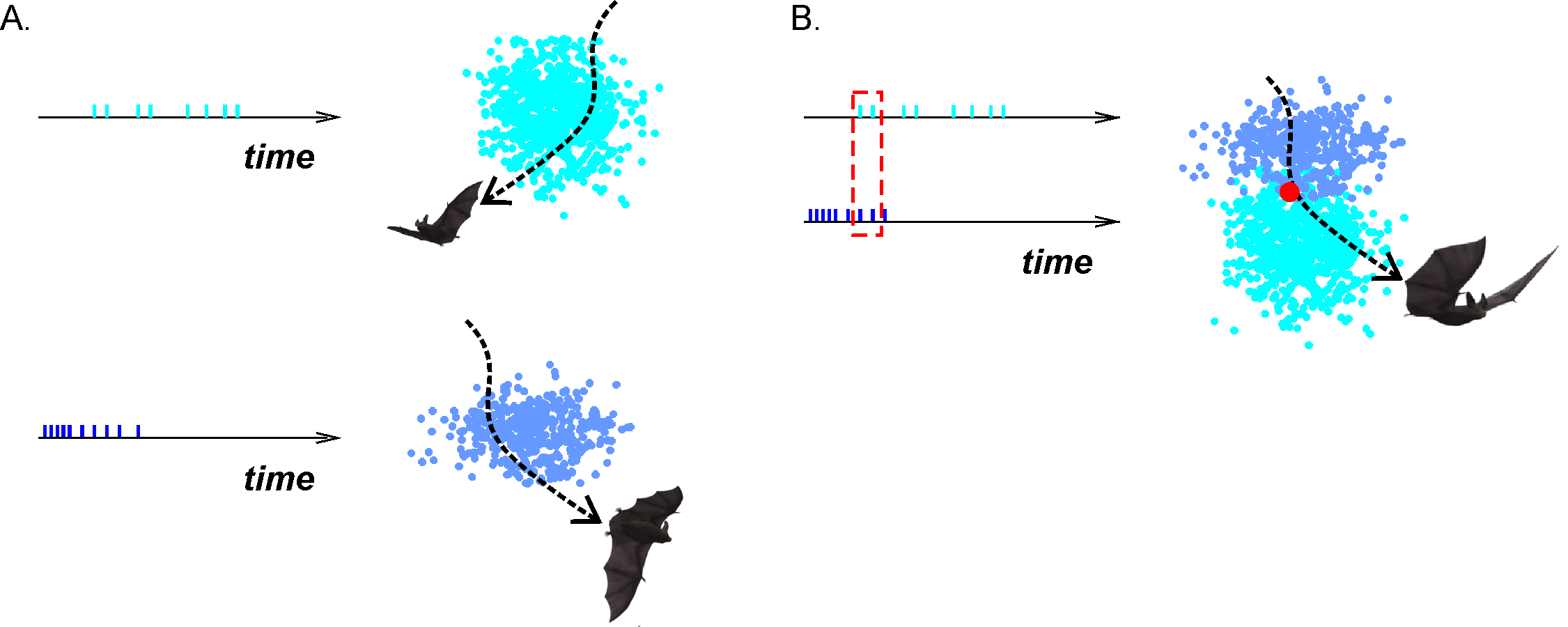}
\caption{\label{SupplFigure1} \textbf{Spatial overlap encoded by temporal coactivity}. (\textbf{A}) Place cells produce spike trains as 
the bat is flying through their respective place fields. (\textbf{B}) Coactivity of two cells (indicated by the dashed box) marks the domain 
where two place fields overlap.} 
\end{figure} 

\begin{figure}[ht] 
\includegraphics[scale=0.84]{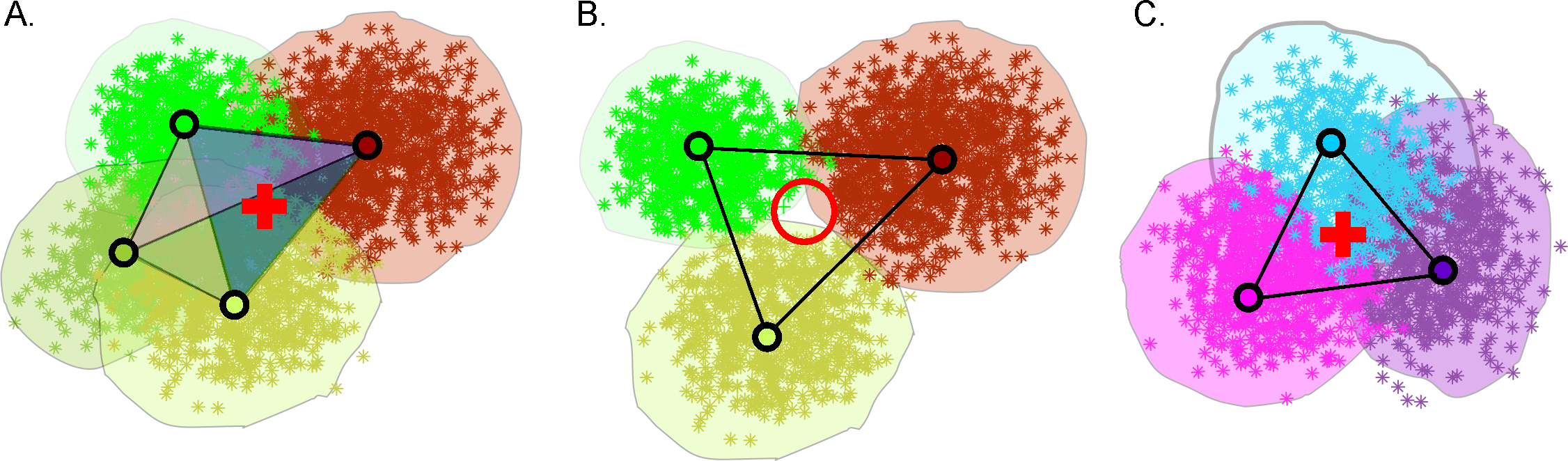}
\caption{\label{SupplFigure1} \textbf{A $2D$ illustration of Helly's theorem}. (\textbf{A}) If every three out of four convex planar regions 
overlap each other, then they necessarily have a common fourth order overlap, marked by the red cross. (\textbf{B}) For three regions, 
pairwise overlappings may not yield a common third order intersection---the hollow spot in the middle is marked by the red circle. 
However, this configuration is statistically rare. (\textbf{C}) Typically, a set of three pairwise overlapping regions is a reliable signature of 
a triple overlap. In other words, a $2D$ clique almost always is a $2D$ simplex (and any simplex is always a clique).} 
\end{figure} 

\begin{figure}[ht] 
\includegraphics[scale=0.9]{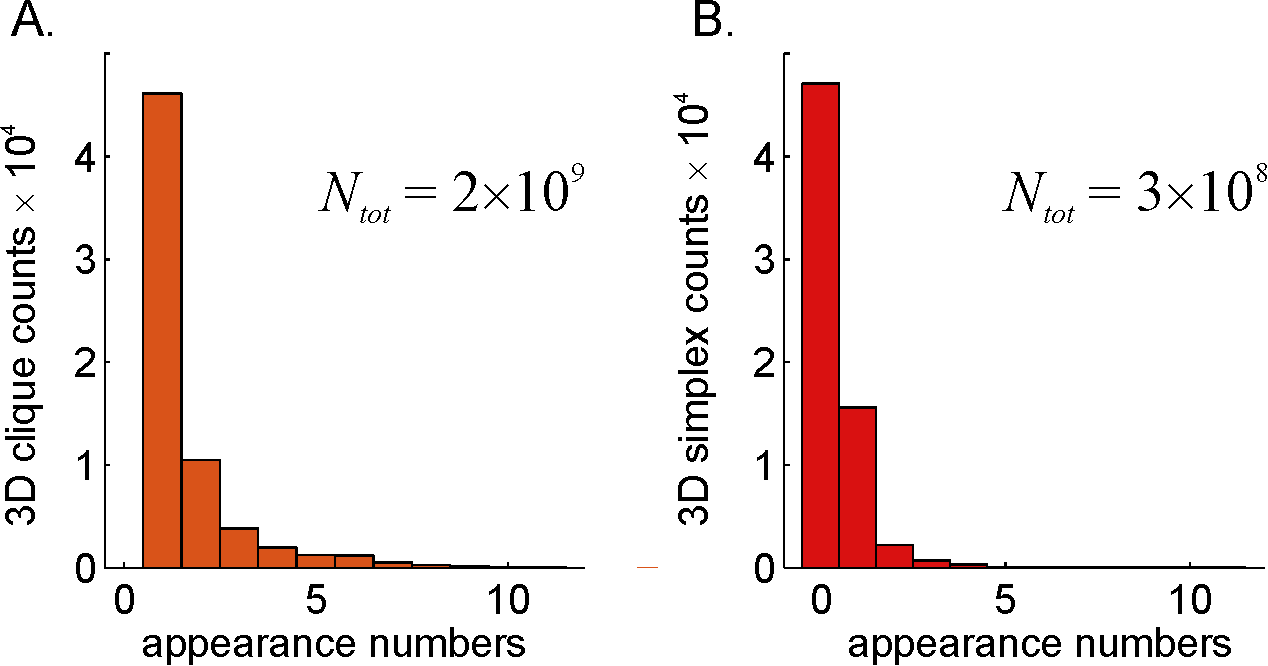}
\caption{\label{SupplFigure2} \textbf{The histograms of appearance numbers of $3D$ cliques}. (\textbf{A}) and $3D$ simplexes (\textbf{B}). Note that the highest occupancy bin on the right panel corresponds to zero appearance rate, i.e., most configurations that observed as cliques, never appear as simplexes, and as a result the total number $N_{tot}$ of cliques and simplexes differ by an order of magnitude. } 
\end{figure} 

\newpage


\newpage

\section{References}

\begin{thebibliography}{99}
\bibitem{OKeefe1} O'Keefe J, Dostrovsky J (1971) The hippocampus as a spatial map. Preliminary evidence from unit activity in the freely-moving rat, \emph{Brain Res}, 34, pp. 171-175.
\bibitem{Best} Best PJ, White AM, Minai A (2001) Spatial processing in the brain: the activity of hippocampal place cells, \emph{Annu Rev Neurosci.}, 24, pp. 459-486.
\bibitem{OKeefe2} O'Keefe J, Nadel L (1978) \emph{The hippocampus as a cognitive map}, New York: Clarendon Press; Oxford University Press. xiv, 570 pp.
\bibitem{Gothard} Gothard KM, Skaggs WE, McNaughton BL (1996) Dynamics of mismatch correction in the hippocampal ensemble code for space: interaction between path integration and environmental cues, \emph{J Neurosci.}, 16, pp. 8027-8040.
\bibitem{Leutgeb} Leutgeb JK, Leutgeb S, Treves A, Meyer R, Barnes CA, et al. (2005) Progressive transformation of hippocampal neuronal representations in "morphed" environments, \emph{Neuron}, 48, pp. 345-358.
\bibitem{Wills} Wills TJ, Lever C, Cacucci F, Burgess N, O'Keefe J (2005) Attractor dynamics in the hippocampal representation of the local environment, \emph{Science}, 308, pp. 873-876.
\bibitem{Touretzky} Touretzky DS, Weisman WE, Fuhs MC, Skaggs WE, Fenton AA, et al. (2005) Deforming the hippocampal map, \emph{Hippocampus}, 15, pp. 41-55.
\bibitem{Diba} Diba K, Buzsaki G (2008) Hippocampal network dynamics constrain the time lag between pyramidal cells across modified environments, \emph{J Neurosci.}, 28, pp. 13448-13456.
\bibitem{eLife} Dabaghian Y, Brandt VL, Frank LM (2014) Reconceiving the hippocampal map as a topological template, \emph{eLife} 10.7554/eLife.03476.
\bibitem{Alvernhe} Alvernhe A, Sargolini F, Poucet B (2012) Rats build and update topological representations through exploration, \emph{Anim. Cogn.}, 15, pp. 359-368.
\bibitem{Poucet} Poucet B, Herrmann T (2001) Exploratory patterns of rats on a complex maze provide evidence for topological coding, \emph{Behav. Processes}, 53, pp. 155-162.
\bibitem{Wu}, Wu X, Foster DJ (2014) Hippocampal Replay Captures the Unique Topological Structure of a Novel Environment, \emph{J Neurosci.}, 34, pp. 6459-6469.
\bibitem{Curto} Curto C, Itskov V (2008) Cell groups reveal structure of stimulus space, \emph{PLoS Comput. Biol.}, 4, e1000205.
\bibitem{Chen1} Chen Z, Gomperts SN, Yamamoto J, Wilson MA (2014) Neural representation of spatial topology in the rodent hippocampus, \emph{Neural Comput.}, 26, pp. 1-39. 
\bibitem{Chen2} Chen Z, Kloosterman F, Brown E, Wilson M (2012) Uncovering spatial topology represented by rat hippocampal population neuronal codes, 
\emph{J Comput. Neurosci.}, 33, pp. 227-255.
\bibitem{Dabaghian} Dabaghian Y, Mémoli F, Frank L, Carlsson G (2012) A Topological Paradigm for Hippocampal Spatial Map Formation Using Persistent Homology, \emph{PLoS Comput. Bio.},l 8: e1002581.
\bibitem{Arai} Arai M, Brandt V, Dabaghian Y (2014) The Effects of Theta Precession on Spatial Learning and Simplicial Complex Dynamics in a Topological Model of the Hippocampal Spatial Map, \emph{PLoS Comput. Biol.}, 10: e1003651. 
\bibitem{Hatcher} Hatcher A (2002) \emph{Algebraic topology}, Cambridge; New York: Cambridge University Press.
\bibitem{Ghrist} Ghrist R (2008) Barcodes: The persistent topology of data, \emph{Bulletin of the American Mathematical Society}, 45, pp. 61-75.
\bibitem{Ulanovsky} Ulanovsky N, Moss CF (2007) Hippocampal cellular and network activity in freely moving echolocating bats, \emph{Nat. Neurosci.}, 10: pp. 224-233.
\bibitem{Yartsev} Yartsev MM, Ulanovsky N (2013) Representation of Three-Dimensional Space in the Hippocampus of Flying Bats, \emph{Science}, 340, pp. 367-372. 
\bibitem{Mizuseki} Mizuseki K, Sirota A, Pastalkova E, Buzsaki G (2009) Theta oscillations provide temporal windows for local circuit computation in the entorhinal-hippocampal loop, \emph{Neuron}, 64, pp. 267-280.
\bibitem{Huhn} Huhn Z, Orbán G, Érdi P, Lengyel M (2005) Theta oscillation-coupled dendritic spiking integrates inputs on a long time scale, \emph{Hippocampus}, 15, pp. 950-962.
\bibitem{Goldman} Goldman-Rakic PS (1995) Cellular basis of working memory, \emph{Neuron} 14: 477-485.
\bibitem{Harris} Harris KD, Csicsvari J, Hirase H, Dragoi G, Buzsaki G (2003) Organization of cell assemblies in the hippocampus, \emph{Nature}, 424, pp. 552-556. 
\bibitem{Buzsaki1} Buzsaki G (2010) Neural syntax: cell assemblies, synapsembles, and readers, \emph{Neuron}, 68, pp. 362-385.
\bibitem{Avis} Avis D, Houle ME (1995) Computational aspects of Helly's theorem and its relatives, \emph{International Journal of Computational Geometry \& Applications}, 05, pp. 357-367.
\bibitem{Babichev} Babichev A, Memoli F, Ji D, Dabaghian Y (2015) Combinatorics of Place Cell Coactivity and Hippocampal Maps. \emph{in submition}; \hyperlink{http://arxiv.org/abs/1509.01677}{arXiv:1509.01677}).
\bibitem{Skaggs} Skaggs WE, McNaughton BL, Wilson MA, Barnes CA (1996) Theta phase precession in hippocampal neuronal populations and the compression of temporal sequences, \emph{Hippocampus}, 6, pp. 149-172.
\bibitem{Buzsaki2} Buzsaki G (2002) Theta oscillations in the hippocampus, \emph{Neuron}, 33, pp. 325-340.
\bibitem{Buzsaki3} Buzsaki G (2005) Theta rhythm of navigation: link between path integration and landmark navigation, episodic and semantic memory, \emph{Hippocampus}, 15, pp. 827-840.
\bibitem{Robbe1} Robbe D, Buzsaki G (2009) Alteration of theta timescale dynamics of hippocampal place cells by a cannabinoid is associated with memory impairment, \emph{J Neurosci.}, 29, pp. 12597-12605.
\bibitem{Robbe2} Robbe D, Montgomery SM, Thome A, Rueda-Orozco PE, McNaughton BL, et al. (2006) Cannabinoids reveal importance of spike timing coordination in hippocampal function, \emph{Nat Neurosci.}, 9, pp. 1526-1533.
\bibitem{Heys} Heys JG, MacLeod KM, Moss CF, Hasselmo ME (2013) Bat and Rat Neurons Differ in Theta-Frequency Resonance Despite Similar Coding of Space, \emph{Science}, 340, pp. 363-367.
\bibitem{London} London M, Häusser M (2005) Dendritic Computation, \emph{Annu Rev. Neurosci.}, 28, pp. 503-532. 
\bibitem{Brody} Brody CD, Romo R, Kepecs A (2003) Basic mechanisms for graded persistent activity: discrete attractors, continuous attractors, and dynamic representations, \emph{Curr. Opin. Neurobiol.}, 13, pp. 204-211.
\bibitem{Magee} Magee JC (2000) Dendritic integration of excitatory synaptic input, \emph{Nat. Rev. Neurosci}, 1, pp. 181-190.
\bibitem{Spruston} Spruston N (2008) Pyramidal neurons: dendritic structure and synaptic integration, \emph{Nat. Rev. Neurosci.}, 9, pp. 206-221.
\bibitem{Rall} Rall W (1989) Cable theory for dendritic neurons. In: Christof K, Idan S, editors. \emph{Methods in neuronal modeling}, MIT Press. pp. 9-92.
\bibitem{Jarsky} Jarsky T, Roxin A, Kath WL, Spruston N (2005) Conditional dendritic spike propagation following distal synaptic activation of hippocampal CA1 pyramidal neurons, \emph{Nat. Neurosci.}, 8, pp. 1667-1676.
\bibitem{Huxter} Huxter JR, Senior TJ, Allen K, Csicsvari J (2008) Theta phase-specific codes for two-dimensional position, trajectory and heading in the hippocampus, \emph{Nat. Neurosci.}, 11, pp. 587-594.
\end{thebibliography}
\end{document}